\documentclass[12pt]{iopart}
\usepackage{amsfonts}
\usepackage{amssymb}
\usepackage{graphicx}
\usepackage{epsfig}
\usepackage{color}
\usepackage{multirow}
\usepackage{booktabs}
\usepackage{array}
\usepackage{xcolor}

\begin{document}

\title{Qualitative quantum simulation of resonant tunneling and localization
with the shallow quantum circuits}

\begin{abstract}
{In a circuit-based quantum computer, the computing is performed via the
discrete-time evolution driven by quantum gates. Accurate simulation of
continuous-time evolution requires a large number of quantum gates and
therefore suffers from more noise. In this paper, we find that shallow
quantum circuits are sufficient to qualitatively observe some typical
quantum phenomena in the continuous-time evolution limit, such as resonant
tunneling and localization phenomena. We study the propagation of a spin
excitation in Trotter circuits with a large step size. The circuits are
formed of two types of two-qubit gates, i.e. XY gates and controlled-$R_{x}$
gates, and single-qubit $R_{z}$ gates. The configuration of the $R_{z}$
gates determines the distribution of the spin excitation at the end of
evolution. We demonstrate the resonant tunneling with up to four steps and
the localization phenomenon with dozens of steps in Trotter circuits. Our
results show that the circuit depth required for qualitative observation of
some typical quantum phenomena is much smaller than that required for
quantitative computation, suggesting that it is feasible to apply
qualitative observations to near-term quantum computers. We also provide a
way to use the physics laws to understand the error propagation in quantum
circuits.}
\end{abstract}

\pacs{11.30.Er, 03.65.Nk, 03.65.-w, 42.82.Et}
\author{}
\maketitle

\author{J. L. Shen, P. Wang{$^{\dagger}$}}
\ead{pwang@ncepu.edu.cn}

\address{Department of Mathematics and Physics, North China Electric Power
University, 102206 Beijing, China}

\noindent\textit{Keywords}: Qualitative quantum simulation, shallow quantum
circuits, resonant tunneling, localization, error propagation

\section{Introduction}

Quantum computing can be used to investigate quantum systems as a universal
simulator \cite{Feynman,Nielsen,Arute}. In quantum mechanics, the time
evolution of quantum states is driven by a Hamiltonian and described by a
unitary operator. In a digital quantum computer, the computing is carried
out by using a set of basic quantum gates, and usually each gate is a
single-qubit or two-qubit unitary operator. The combination of these basic
gates allows us to implement the evolution operator of a multi-qubit system.
A specific approach is Trotter-Suzuki decomposition \cite%
{Childs2021,AGuzik2024}, in which we approximate the continuous-time
evolution with a discrete-time evolution. For a general local-interaction
Hamiltonian, we can explicitly construct the evolution operator for a short
time, i.e. one time step, from quantum gates. By repetitive gates of one
time step for $N_{T}$ times, we realize the target time evolution. With a
smaller step size, the discrete-time evolution is closer to continuous-time
evolution, however, this requires a larger $N_{T}$, i.e. more quantum gates.
Considering a practical device \cite{Preskill}, quantum computing is
inaccurate due to decoherence and imperfect control, and usually the error
increases with the gate number \cite{Landauer,Unruh,Raimond}. Fault-tolerant
quantum computing using quantum error correction is able to remove the error
but challenging using today's technologies, because of the large qubit
overhand for encoding \cite{Shor,Google2025}. A family of practical methods
have been developed to mitigate errors, however the gate number is usually
limited due to the finite error rate on the physical level \cite%
{LYa,YuanX,Cai2023}. Therefore we can only realize the discrete-time Trotter
evolution with a small number of Trotter steps. This motivates researches on
the effect of large step size, i.e. few Trotter steps. Trotter errors
induced by large step sizes in digital quantum simulation have received
extensive attention \cite{Malley,SZhu,Zeng2025}. Some studies show that the
Trotter step sizes can separate quantum chaotic phase from localized phase
and comparatively large Trotter steps can retain controlled errors for local
observables \cite{ZollerSA,Zoller}.

In this paper, we are interested in, when the step size is large (or
equivalently the steps are few), whether some typical physical effects in
the limit of continuous-time can still be observed. The physical effects
that we focus on are the resonant tunneling phenomenon and the localization
in disordered systems \cite{DJ,Segev2013}. In the systems
considered in this work, we find that, in a large-step-size Trotter
circuit, the resonant tunneling with $n$ resonant peaks can be observed in
circuits with $n+1$ Trotter steps. Experiments on an IBM quantum computer
are implemented to demonstrate the resonant tunneling with up to three
peaks. We also study the spin transport with the disordered configurations
of the $R_{z}$ gates (we will specify these gates later) using the large
step size. The numerical simulation of circuits with $15$ qubits and tens of
Trotter steps exhibits the localization in the disordered configuration. The
results indicate that shallow quantum circuits on near-term quantum
computers are sufficient to qualitatively simulate some typical physical
phenomena. The localization phenomenon of the spin excitation distribution
implies that the bit-flip error does not affect the measurement on distant
qubits if the configuration of the $R_{z}$ gates is disordered. These
conclusions can be generalized if we replace XY gates with controlled-$R_{x}$
gates, which can transform one spin excitation into multiple spin
excitations.

This paper is organized as follows. In Sec. \ref{Model}, we discuss the
quantum transverse-field XY model and corresponding quantum circuits, the
map between the Hamiltonian of model and corresponding circuit is
established in the limit of small Trotter step size. In Sec. \ref{Resonant
tunneling for large Trotter step size}, we discuss the propagation of the
spin excitation, and compare resonant tunneling effects in circuits in the
small-step-size limit and the large-step-size limit. In Sec. \ref%
{Localization for large Trotter step size}, we investigate the transport of
the spin excitation in the ordered and disordered configurations of
single-qubit $R_{z}$ gates. Conclusion is given at the end of the paper.
\begin{figure}[t]
\centering
\includegraphics[bb=0 0 658 503, width=12 cm, clip]{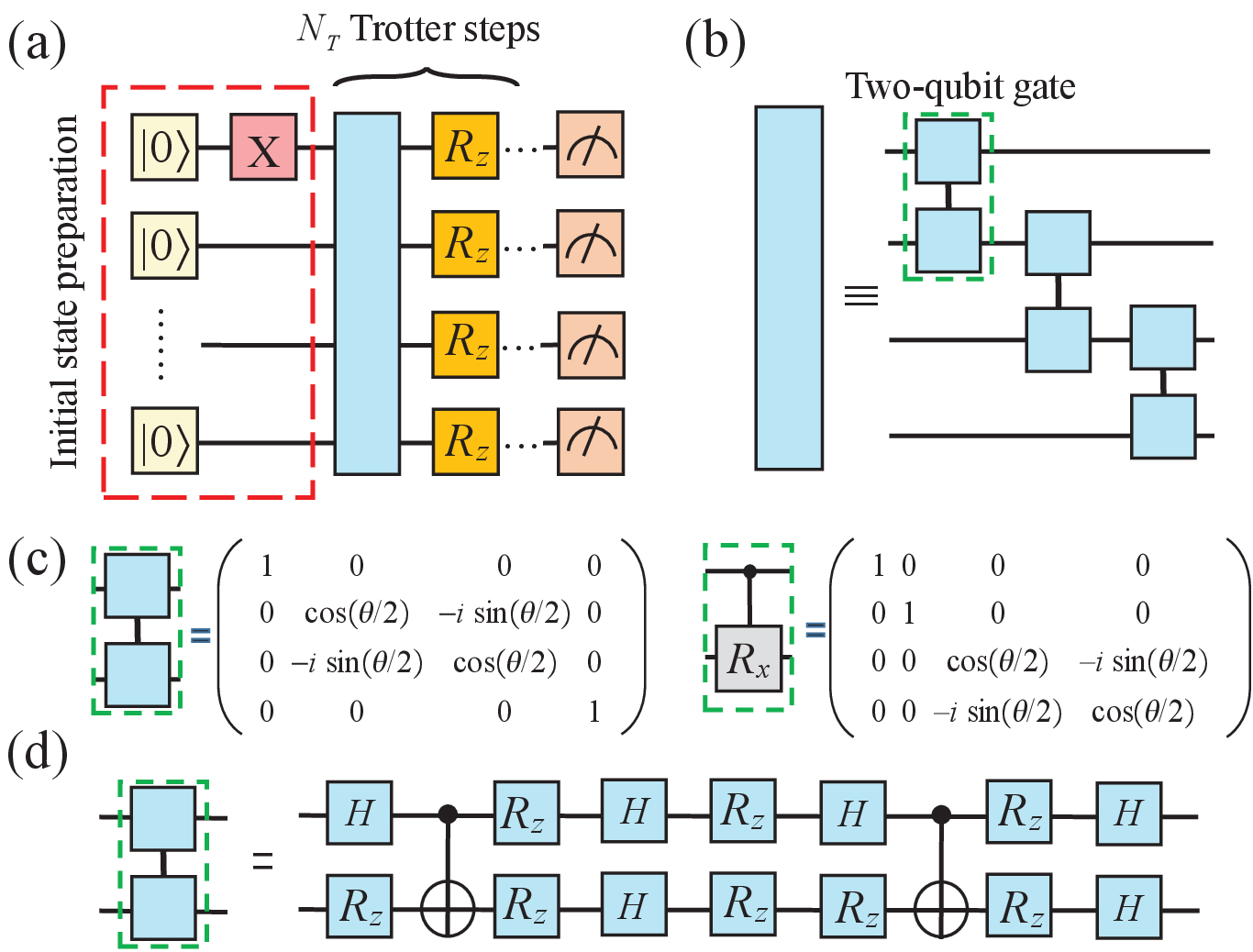} %
\caption{{} (Color online) (a) The schematic diagram of quantum circuit,
which includes the initialization, $N_{T}$ Trotter steps, and measurements.
The preparation for the initial state is in the red dashed box including a
NOT gate (i.e. the X gate). The blue rectangle represents a layer of
two-qubit gates. The orange rectangles represent single-qubit $R_{z}$ gates.
(b) The schematic diagram of a layer of two-qubit gates. (c) The matrix
representations for two types of 2-qubit gates: XY gates and controlled-$%
R_{x}$ gates. (d) The XY-gates are decomposed into a sequence of
elementary gates native to the quantum hardware.}
\label{fig1}
\end{figure}

\section{Model}

\label{Model}

In this paper, we study the particle transport in the discrete-time
evolution in the quantum transverse-field XY model. The purpose of this
study is to investigate whether some typical quantum phenomena occurring in
the continuous-time limit can be qualitatively observed in discrete-time
evolution when the step size is large. The typical quantum phenomena we
concerned here including resonant tunneling and localization effect, which
caused by interference during the particle transport. Generally, the quantum
transverse-field XY model can be used to describe the propagation of spin
excitations, or particle transport when spin excitations can be treated as
particles \cite{JW}. Additionally, in the quantum transverse-field XY model,
the corresponding discrete-time evolution operator can be mapped into a
quantum circuit according to the Trotter-Suzuki decomposition. Therefore, we
investigate the particle transport with the quantum transverse-field XY model%
\begin{equation}
H=\sum_{j}^{N-1}J_{j}H_{j,j+1}^{XY}+\sum_{j}^{N}V_{j}H_{j}^{Z},
\label{TransXY}
\end{equation}%
where $j$ is the lattice site, $N$ is the system size, $J_{j}$ is the
interaction strength and $V_{j}$ is the transverse-field strength. $%
H_{j,j+1}^{XY}=(\sigma _{j}^{x}\sigma _{j+1}^{x}+\sigma _{j}^{y}\sigma
_{j+1}^{y})/4$ and $H_{j}^{Z}=\sigma _{j}^{z}/2$, where $\sigma _{j}^{i}$ $%
(i=x,y,z)$ represents the Pauli matrix at the $j$th site. In the case of $%
J_{j}<0$ and $V_{j}<0$, the ground state of the transverse-field XY chain is
$\left\vert 00...0\right\rangle $. In this work, we consider the time
evolution of the initial state $\left\vert \psi (0)\right\rangle =\left\vert
10...0\right\rangle $ which represents a spin excitation on the first site. The time evolution is restricted to the single-spin-excitation
subspace. Therefore, the probability of measuring the $i$th qubit in the
state $\left\vert 1\right\rangle $ can be obtained directly from standard
projective measurements.

The time evolution operator $U(t)$ of quantum transverse-field XY model can
be approximated with a quantum circuit. According to Trotter-Suzuki
decomposition, $U(t)$ can be expanded approximately%
\begin{equation}
U(t)=e^{-iHt}\approx \lbrack \prod_{j=1}^{N-1}U_{j,j+1}^{XY}(J_{j}\tau
)\prod_{j=1}^{N}U_{j}^{Z}(V_{j}\tau )]^{N_{T}},  \label{TS}
\end{equation}%
where $N_{T}$ is the number of Trotter steps, $\tau =t/N_{T}$ is the size of
each Trotter step. $U_{j,j+1}^{XY}(\theta _{j})=e^{-iH_{j,j+1}^{XY}\theta
_{j}}$ and $U_{j}^{Z}\left( \phi _{j}\right) =e^{-iH_{j}^{Z}\phi _{j}}$,
where $\theta _{j}=J_{j}\tau $ and $\phi _{j}=V_{j}\tau $.
\begin{figure*}[t]
\centering
\includegraphics[bb=0 0 1890 715, width=16 cm, clip]{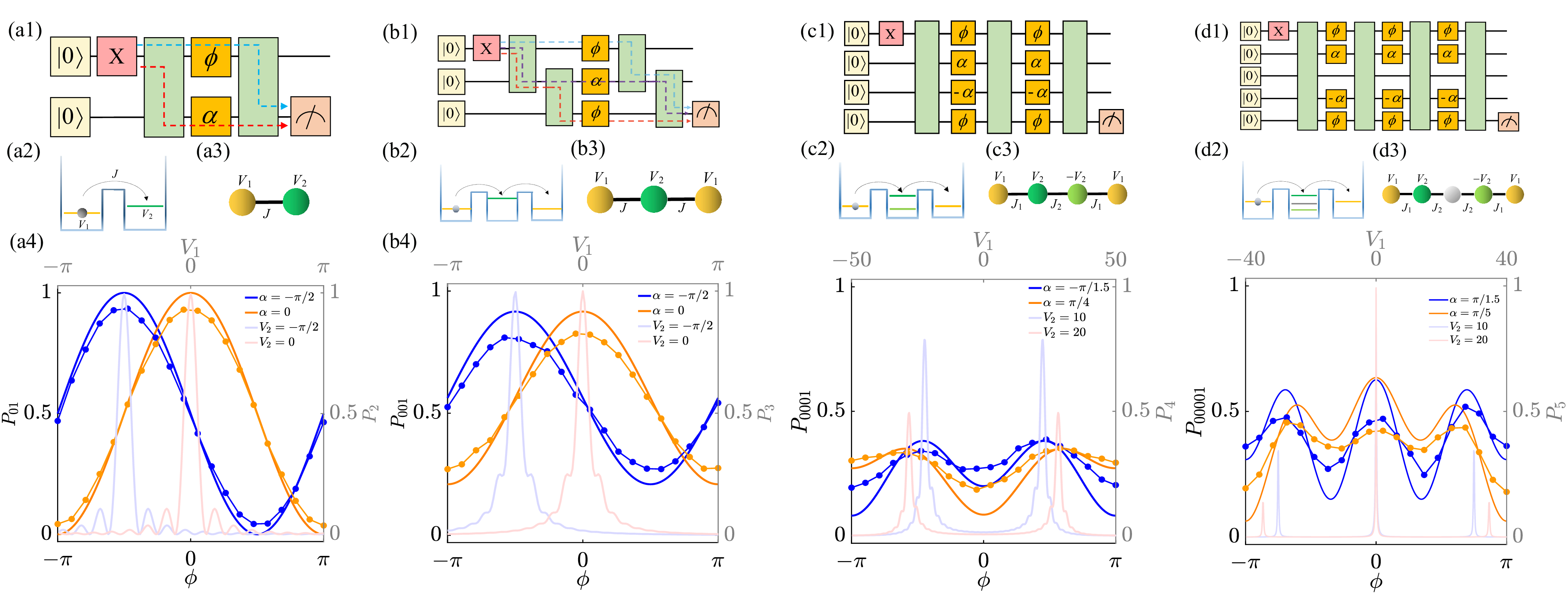}
\caption{ (Color online) (a1)-(d1) The circuit systems. The parameters of $%
R_{z}$ gates are marked on the orange squares. For convenience, in (c1) and
(d1) we use single green rectangle to denote a layer of two-qubit gates.
(a2)-(d2) The quantum wells. The energy levels of the wells are denoted by
the horizontal lines in the wells. (a3)-(d3) The tight-binding chain.
(a4)-(d4) Numerical result of the discrete-time evolution. The solid curves
represent the probability of observing the spin excitation on the last qubit
after the time evolution. The curves with circles are the results obtaining
on a IBM quantum computer. The curves with light colors are the numerical
result of the continuous-time evolution.}
\label{fig2}
\end{figure*}
The corresponding discrete-time evolution is realized with the quantum
circuit as shown in Fig. \ref{fig1}(a). The time evolution of each term,
i.e. $U^{XY}$ and $U^{Z}$, are two-qubit XY gate and single-qubit $R_{z}$
gate, respectively. The circuit has $N$ qubits $\{q_{1},q_{2},...,q_{N}\}$
and $N_{T}$ Trotter steps, and every Trotter step contains one layer of
two-qubit XY\ gates and one layer of single-qubit $R_{z}$ gates (We neglect $%
R_{z}$ gates in the last Trotter step, because these gates does not have any
effect on the distribution of the spin excitation). In a quantum circuit, a
NOT gate (i.e. the Pauli $\sigma _{x}$ matrix) can flip the qubit $%
\left\vert 0\right\rangle $ to $\left\vert 1\right\rangle $, therefore we
prepare the initial state $\left\vert \psi (0)\right\rangle $ by applying a
NOT gate on the first qubit, i.e. $\left\vert \psi (0)\right\rangle =\sigma
_{1}^{x}\left\vert 0\right\rangle ^{N}=\left\vert 10...0\right\rangle $ [see
the dashed rectangle in Fig. \ref{fig1}(a)]. The matrix representation of XY
gate is shown in Fig. \ref{fig1}(c). Note that the XY gate
cannot be executed as a whole in a quantum circuit. It must be decomposed
into a sequence of elementary gates native to the quantum hardware. The
decomposition is not unique; in this work, we implement it as the sequence $%
R_{i}^{z}(\pi
/2),H_{j},C_{j}(X_{i}),R_{i}^{z}(\pi/2),H_{i},H_{j},R_{i}^{z}(\theta
/2),R_{j}^{z}(-\theta /2),R_{i}^{z}(\pi/2),H_{j}$, $C_{j}(X_{i}),R_{i}^{z}(-%
\pi /2),R_{j}^{z}(\pi /2),H_{i},H_{j}$, where $R_{i}^{z}$ ($H_{i}$)
represents a rotation-Z (Hadamard gate) on the $i$th qubit , $C_{j}(X_{i})$
represents a controlled-X (CNOT) gate with control qubit $i$ and target
qubit $j$. Thus, the two-qubit XY gates shown Fig. \ref{fig1}(b) corresponds
in practice to the gate sequence displayed in Fig. \ref{fig1}(d).

The behavior of a spin excitation under discrete-time evolution depends on
step size. When the Trotter step size is sufficiently small, the propagation
of a spin excitation in quantum circuit is equivalent to the particle
transport in continuous-time evolution. In this case, the propagation of
spin excitation can exhibit some typical physical phenomena in particle
transport. A natural question to ask is, in shallow quantum circuits with a
large Trotter step size, whether some physical phenomena during
continuous-time evolution nevertheless remains, so that we can observe these
phenomena using fewer quantum gates. In the following text, we show that we
observe the resonant tunneling and localization effect in shallow circuits
even if the Trotter step size is large, similar to those
observed in the small-step-size limit.

\section{Resonant tunneling for large Trotter step size}

\label{Resonant tunneling for large Trotter step size}

In this section, we study the resonance phenomenon associated with the
transport of a spin excitation during discrete-time evolution with a large
step size. We consider systems with $N=2$ and $N=5$ (the cases $%
N=3$ and $N=4$ are provided in the appendix). In each case, the spin
excitation is initialized by a NOT gate and transmitted via the XY gate. For
the systems under consideration, an expression is derived for the
probability of observing a spin excitation on the last qubit. The resonance
phenomena exhibited by the shallow quantum circuits are found to closely
resemble those of the continuous-time limit in terms of certain features,
including the number of peaks and the factors that determine their positions.%

We also study the transport of a spin excitation through the controlled-$%
R_{x}$ gate. In quantum computing, computational errors occur due to the
imperfect control and decoherence always error. The typical one is bit-flip
error corresponding to an unwanted NOT gate. So studying the transmission of
the spin excitation can help us understand the propagating of bit-flip
errors. In actual quantum circuits, errors are not only transmitted, but
also replicated. For example, a single-qubit error will become a multi-qubit
error after passing through the controlled-$R_{x}$ gate. This effect
potentially has a greater impact in quantum computing, because multi-qubit
errors can lead to the failure of quantum error correction. So in this
section, we also study the behavior of a spin excitation propagating through
the controlled-$R_{x}$ gate and find the resonance phenomenon analogous to
that propagating through the XY gate.

\subsection{Resonant tunneling in two-qubit circuit}

\label{Resonant tunneling in two-qubit circuit}

This subsection discusses the case of $N=2$. The Hamiltonian of the
transverse-field XY model with two sites is%
\begin{equation}
H=J(\sigma _{1}^{x}\sigma _{2}^{x}+\sigma _{1}^{y}\sigma
_{2}^{y})/4+V_{1}\sigma _{1}^{z}/2+V_{2}\sigma _{2}^{z}/2.  \label{XY2}
\end{equation}%
According to the Trotter-Suzuki decomposition, the time evolution operator
can be approximated using a sequence of quantum gates%
\begin{equation}
U(t)=e^{-iHt}\approx \lbrack U_{1,2}^{XY}(J\tau )U_{1}^{Z}(V_{1}\tau
)U_{2}^{Z}(V_{2}\tau )]^{N_{T}},  \label{Ut2}
\end{equation}%
where $U_{j,j+1}^{XY}(J_{j}\tau )=e^{-i(\sigma _{j}^{x}\sigma
_{j+1}^{x}+\sigma _{j}^{y}\sigma _{j+1}^{y})J_{j}\tau /4}$, $U_{j}^{Z}\left(
V_{j}\tau \right) =e^{-i\sigma _{1}^{z}V_{j}\tau /2}$, and $\tau =t/N_{T}$.
The right side of "$\approx $"\ in Eq. {(\ref{Ut2}) represents the
discrete-time evolution. }The accuracy of the approximation increases with $%
N_{T}$. In Fig. \ref{fig2}(a1), we show the schematic diagram of the quantum
circuit corresponding to {discrete-time evolution when }$N_{T}=2$ (the last
two $R_{z}$ gates have been ignored, as in the other systems). As we can
see, the yellow squares represent that the qubits are initialized to $%
\left\vert 00\right\rangle ,$ which is the ground state of $H$ in the case
of $V_{1},V_{2}<0$ and $J\ll V_{1}+V_{2}$. Following the initialization, a
NOT gate on the first qubit flips $\left\vert 00\right\rangle $ into $%
\left\vert 10\right\rangle $, which represents that there is a spin
excitation on the first qubit. $\left\vert 10\right\rangle $ is the initial
state that we want to prepare. Due to the symmetry of Hamiltonian, the spin
excitation lies in the subspace spanned by \{$\left\vert 10\right\rangle $, $%
\left\vert 01\right\rangle $\}. In this single-spin-excitation subspace, we
can regard a spin excitation as a particle moving in a $2$-site
tight-binding chain [see Fig. \ref{fig2}(a3)], and the corresponding
Hamiltonian is%
\begin{equation}
H=J\left( \left\vert 10\right\rangle \left\langle 01\right\vert +\left\vert
01\right\rangle \left\langle 10\right\vert \right) +V_{1}\left\vert
10\right\rangle \left\langle 10\right\vert +V_{2}\left\vert 01\right\rangle
\left\langle 01\right\vert ,  \label{H2}
\end{equation}%
where $\left\vert 10\right\rangle $ or $\left\vert 01\right\rangle $
represents a particle in the first or second site respectively, $J$ is the
tunneling strength, $V_{1}$ and $V_{2}$ are the on-site potentials. In this
chain system with fixed parameters $J$, $V_{2}$ and variable $V_{1}$, a
resonance phenomenon can be observed \cite{Aidelsburger,Tsu,Peck}: assuming
the particle is on the first site at $t=0$, the probability of finding it on
the second site at time $t,$ denoted by $P_{2}(V_{1},t),$ reaches its
maximum when $V_{1}=V_{2}$. We numerically simulate this phenomenon in a
discrete-time evolution with a large $N_{T}$ in Fig. \ref{fig2}(a4).\ We
plot $P_{2}$ at $t=15$ (in units of $1/J$,{\ }$J=0.1$) as transparent lines
for $V_{2}=0,-\pi /2$. As expected, $P_{2}$ exhibits a single resonance peak
at $V_{1}=V_{2}$. We can understand this phenomenon more intuitively with
the help of a double-well system, as shown in Fig. \ref{fig2}(a2). Supposing
a particle is initially confined in the left well, it will tunnel to the
right well with a certain probability. When the potential energies on both
sides are equivalent (i.e. $V_{1}=V_{2}$), the tunneling probability reaches
its maximum.

An important consideration is whether a resonant effect, similar to that
observed in the large $N_{T}$ limit, can be qualitatively observed when $%
N_{T}$ is small. To explore this, we now consider the case of $N_{T}=2$. The
parameters are redefined as $\theta =J\tau ,\phi =V_{1}\tau ,$ and $\alpha
=V_{2}\tau $. Therefore, scanning the circuit parameter $\phi$
corresponds to scanning $V_{1}$ in the continuous-time evolution, and
similarly for the other systems studied. Our concern is the probability of
finding spin excitation on the $2$nd qubit. Figure\textbf{\ }\ref{fig2}(a1)
shows two propagation paths of the spin excitation from the $1$st to the $2$%
nd qubit. The blue path contributes $-i\sin (\theta /2)\cos
(\theta /2)e^{i(\alpha -\phi )/2}$ to the amplitude, the purple path
contributes $-i\sin (\theta /2)\cos (\theta /2)e^{-i(\alpha
-\phi )/2}$ to the amplitude, and so the final state of the quantum circuit
reads%
\begin{equation}
U_{1,2}^{XY}\left( \theta \right) U_{2}^{Z}\left( \alpha \right)
U_{1}^{Z}\left( \phi \right) U_{1,2}^{XY}(\theta )\left\vert 10\right\rangle
=A_{10}\left\vert 10\right\rangle +A_{01}\left\vert 01\right\rangle ,
\nonumber
\end{equation}%
where $A_{01}=-i\sin (\theta /2)\cos (\theta /2)(e^{-i(\alpha
-\phi )/2}+e^{i(\alpha -\phi )/2})$. The probability of the spin excitation
measured on the second qubit is%
\begin{equation}
P_{01}(\theta ,\phi )=2\sin ^{2}(\theta /2)\cos ^{2}(\theta
/2)\left( 1+\cos (\alpha -\phi )\right).  \label{P01}
\end{equation}%
In Fig. \ref{fig2}(a4), we plot $P_{01}$ as a function of $\phi $ with $%
\theta =\pi /2$ and $\alpha =0,-\pi /2$. The accurate results (solid lines)
computed using QuESTlink coincide with the experimental outcomes (solid
lines with point symbols) computed using the IBM quantum device
"ibmq\_rome". This device is a 5-qubit processor with qubits
connected in a linear chain. For context, its single-qubit gate errors are
approximately ${0.03\%}${, two-qubit gate errors range from }$0.7\%${\ to }$%
0.9\%${, and readout errors are between }${2\%}${\ and }${4\%}$. The
resonance observed in circuits with two Trotter steps exhibits two
characteristics that are consistent with those in the continuous-time limit.
On the one hand, Eq. {(\ref{P01}) }indicates that $P_{01}$ exhibits a single
resonance peak, consistent with the continuous-time limit. On the other
hand, the peak occurs at $\phi =\alpha ,$ where $\phi $ and $\alpha $
corresponds to $V_{1}$ and $V_{2}$ in the continuous-time case.

\subsection{Resonant tunneling five-qubit systemr}

\label{Resonant tunneling five-qubit system}

Then we discuss the five-qubit system. The Hamiltonian of the $5$-site
transverse field XY chain we studied is%
\begin{eqnarray}
H &=&J_{1}\sum_{j=1,4}(\sigma _{j}^{x}\sigma _{j+1}^{x}+\sigma
_{j}^{y}\sigma _{j+1}^{y})/4+J_{2}\sum_{j=2,3}(\sigma _{j}^{x}\sigma
_{j+1}^{x}+\sigma _{j}^{y}\sigma _{j+1}^{y})/4 \\
&&+V_{1}\sigma _{1}^{z}/2+V_{2}\sigma _{2}^{z}/2-V_{2}\sigma
_{4}^{z}/2+V_{1}\sigma _{5}^{z}/2.  \nonumber
\end{eqnarray}%
The quantum circuit implementing the discrete-time evolution is shown in Fig.%
\textit{\ }\ref{fig2}(d1). In the single-particle subspace, the equivalent $%
5 $-site tight-binding chain is%
\begin{eqnarray}
H &=&[J_{1}\left( \left\vert 10000\right\rangle \left\langle
01000\right\vert +\left\vert 00010\right\rangle \left\langle
00001\right\vert \right)  \nonumber \\
&&+J_{2}(\left\vert 01000\right\rangle \left\langle 00100\right\vert
+\left\vert 00100\right\rangle \left\langle 00010\right\vert )+\mathrm{h.c.}]
\nonumber \\
&&+V_{1}\left\vert 10000\right\rangle \left\langle 10000\right\vert
+V_{2}\left\vert 01000\right\rangle \left\langle 01000\right\vert  \label{M5}
\\
&&-V_{2}\left\vert 00010\right\rangle \left\langle 00010\right\vert
+V_{1}\left\vert 00001\right\rangle \left\langle 0001\right\vert .  \nonumber
\end{eqnarray}%
The schematic diagram of the Hamiltonian Eq. ({\ref{M5}}) is shown in Fig. %
\ref{fig2}(c4). Let $P_{5}(V_{1},t)$ denote the probability of finding the
particle on the $5$th site. Assuming {$J_{1}\ll J_{2},V_{2}$, then the
strong coupling among the three central sites effectively binds them into a
composite quantum well. The energy levels of the central well are $0,\pm
\sqrt{2J_{2}^{2}+V_{2}^{2}}$ (the contribution of $J_{1}$ is negligible). As
the on-site potential $V_{1}$ at the edge sites is varied, resonant
tunneling occurs when $V_{1}$ matches with one of these energy levels,
leading to a maximum in the occupation probability on the fifth site. This
gives rise to three distinct resonance peaks in $P_{5}(V_{1},t).$ The
central peak remains fixed at $V_{1}=0,$ independent of the system
parameters. In contrast, the positions of the two outer peaks are $\pm \sqrt{%
2J_{2}^{2}+V_{2}^{2}}$, which is governed by $J_{2}$ and $V_{2}.$ }We
numerically simulate $P_{5}$ (see the transparent lines in Fig. \ref{fig2}%
(d4)) with system parameters $J_{1}=0.1$, $t=40$ (in units of $1/J_{1}$)$,$ $%
J_{2}=20$, and $V_{2}=10,20$. These simulations demonstrate that $P_{5}$
exhibits the above phenomenon.

When $N_{T}$ is small, we find that only four Trotter steps are required for
the circuit to qualitatively exhibit resonant tunneling. We investigate the
propagation of the spin excitation in the circuit shown in Fig. \ref{fig2}%
(d1). We redefine parameters as $J_{1}\tau =\theta _{1},J_{2}\tau =\theta
_{2},V_{1}\tau =\phi ,V_{2}\tau =\alpha ,$ and denote the probability of
finding spin excitation on the $5$th qubit as $P_{00001}$.\
According to Eq. ({\ref{Papp}}) in Appendix \ref{The amplitude of the paths}%
, by enumerating and listing the amplitudes of all paths, the expression for
$P_{00001}$ can be obtained,%
\begin{equation}
P_{00001}=\sum_{m=0}^{3}g_{m}\cos m\phi  \label{P5}
\end{equation}%
where $g_{0}=\sum_{m=0}^{3}f_{m}^{2},g_{1}=2\left(
f_{2}f_{3}+f_{1}f_{2}+f_{1}f_{0}\right) ,g_{2}=2\left(
f_{0}f_{2}+f_{3}f_{1}\right) ,g_{3}=2f_{0}f_{3},$ and%
\begin{equation}
\left\{
\begin{array}{c}
f_{3}=\sin ^{2}(\theta _{1}/2)\sin ^{2}(\theta _{2}/2)(18+32\cos \alpha \cos
(\theta _{1}/2)\cos ^{3}(\theta _{2}/2)\times \\
(-1+3\cos \theta _{2})+\cos \theta _{2}(15+16\cos \theta _{1}\cos \theta
_{2})+14\cos (2\theta _{2}) \\
+4\cos (2\alpha )(1+\cos \theta _{1})(1+16\cos \alpha \cos (\theta
_{1}/2)\cos ^{3}(\theta _{2}/2) \\
+4\cos \theta _{2}+3\cos (2\theta _{2}))+\cos (3\theta _{2}))/32 \\
f_{2}=\cos (\theta _{2}/2)\sin ^{2}(\theta _{1}/2)\sin ^{2}(\theta
_{2}/2)(2\cos \alpha \cos \theta _{2}(1+3\cos \theta _{1}) \\
+\cos (\theta _{1}/2)((3+(4+8\cos 2\alpha )\cos \theta _{1})\cos (\theta
_{2}/2)+\cos (3\theta _{2}/2)))/2 \\
f_{1}=3\cos (\theta _{1}/2)\cos (\theta _{2}/2)\sin ^{2}(\theta _{1}/2)\sin
^{2}(\theta _{2}/2)(2\cos \alpha \cos \theta _{1} \\
+\cos (\theta _{1}/2)\cos (\theta _{2}/2)) \\
f_{0}=\cos (\theta _{1}/2)\sin ^{2}(\theta _{2}/2)\sin ^{2}\theta _{1}%
\end{array}%
\right. .
\end{equation}%
Using polynomial analysis tools, it is straightforward to find that the
positon of peak is located at $\phi =0,\pm \arccos t_{0}$ where $%
t_{0}=(-4g_{2}+\sqrt{16g_{2}^{2}-48g_{3}g_{1}+144g_{3}^{2}})/(24g_{3})$.
That is, under the condition that $|t_{0}|<1$, the resonance observed in
circuits with four Trotter steps exhibits two characteristics that are
consistent with those in the continuous-time limit. First, the number of
peaks is three in both cases. Second, the central peak remains fixed at $%
\phi =0$, mirroring the behavior in the continuous-time case; the positions
of the two outer peaks depend on $\alpha $ and $\theta _{2}$, also in
agreement with the continuous-time evolution, where the outer peak locations
are governed by $V_{2}$ and $J_{2}.$ Figure \ref{fig2}(d4) exhibits $%
P_{00001}$ (the solid lines) as a function of $\phi $ by fixing $\alpha
,\theta _{1}=\pi /3$ and $\theta _{2}=\pi /1.2$. We compare two cases that $%
\alpha $ is $\pi /1.5,\pi /5$ respectively. The distance between the peaks
is changed when $\alpha $ is adjusted, which is the characteristic of the
resonant tunneling effect.

In addition to the $N=2$ and $N=5$ cases discussed above, we
also investigate systems with $N=3$ and $N=4.$ The detailed calculations for
these cases are provided in the Appendix, and the results for all system
sizes are summarized in table \ref{comparison}. The code for the shallow
quantum circuits and numerical evolution is available on GitHub at:
https://github.com/nkuwangpeng-hue/QualitativeQuantumSimulation/tree/master.
\begin{table}[h]
\caption{Comparison of phenomena in the continuous-time
evolution limit and the shallow quantum circuit, where $\protect\alpha $
corresponds to $V_{2}$ and $\protect\theta _{2}$ to $J_{2}$}
\label{comparison}\centering
\begin{tabular}{|c|c|c|c|c|}
\hline
\multirow{2}{*}{\(N\)} & \multicolumn{2}{c|}{Continuous-time evolution limit}
& \multicolumn{2}{c|}{Shallow circuit} \\ \cline{2-5}
& Peaks & Dependence & Peaks & Dependence \\ \hline
2 & 1 & $V_2$ & 1 & $\alpha$ \\ \hline
3 & 1 & $V_2$ & 1 & $\alpha$ \\ \hline
4 & 2 & $J_2, V_2$ & 2 & $\theta_2, \alpha$ \\ \hline
5 & 3 & two: $J_2, V_2$; one: $0$ & 3 & two: $\theta_2, \alpha$; one: $0$ \\
\hline
\end{tabular}%
\end{table}

\subsection{Discussion}

\label{Discussion} The results above show that shallow quantum
circuits with $N_{T}=n+1$ layers (where $n$ is the number of peaks in the
continuous-time evolution limit) capture certain resonance features of that
limit. This section provides a theoretical explanation for this observation.

The probability of detecting an error on the final qubit can be expressed in
a general form. The error (i.e., a spin excitation) can reach the final
qubit along any of several possible paths. A path is uniquely represented by
the sequence \{$k_{i}$\} with $k_{i}$ denoting the qubit position at which
the error resides at the end of the $i$-th layer. From the action of the
two-qubit and single-qubit gates on the quantum state $\sigma
_{i}^{x}\left\vert 0\right\rangle ^{N}$, we can derive the expression for
the amplitude of the path \{$k_{i}$\}%
\begin{equation}
A(\{k_{i}\})=a(\{k_{i}\})e^{-iS/2}e^{i\sum_{i=1}^{N_{T}-1}\phi _{k_{i}}}
\label{A}
\end{equation}%
and the probability of detecting the error on the last qubit%
\begin{equation}
P_{0\cdot \cdot \cdot 1}=\sum_{\{k_{i}\},\{k_{i}^{\prime
}\}}a(\{k_{i}\})a^{\ast }(\{k_{i}^{\prime }\})e^{i\sum_{i=1}^{N_{T}-1}\phi
_{k_{i}}}e^{-i\sum_{i=1}^{N_{T}-1}\phi _{k_{i}^{\prime }}},  \label{P}
\end{equation}%
where $\theta _{j}=${\ }$J_{j}\tau ,$ $\phi _{j}=V_{j}\tau $, $%
a(\{k_{i}\})=\prod\limits_{j}\cos ^{c_{j}}(\theta
_{j}/2)\prod\limits_{j^{\prime }}(-i\sin (\theta _{j^{\prime
}}/2))^{c_{j^{\prime }}},S=(N_{T}-1)\sum_{j=1}^{n}\phi _{j}$. The factor $%
\prod\limits_{j}\cos ^{c_{j}}(\theta _{j}/2)$ (resp. $(\prod\limits_{j^{%
\prime }}(-i\sin (\theta _{j^{\prime }}/2))^{c_{j^{\prime }}}$) originates
from XY gates along the path \{$k_{i}$\} that involve no (resp. do) qubit
transfer, where $c_{j}$ (resp. $c_{j^{\prime }}$) denotes the count of XY$%
_{j,j+1}$ gates satisfying the respective condition, and the factor $%
e^{-iS/2}e^{i\sum_{i=1}^{N_{T}-1}\phi _{k_{i}}}$ originates from the error
traversing the $N_{T}-1$ layers of $R_{z}$ gates. The detailed calculation
is shown in Appendix.

The Hamiltonian under consideration determines the minimum number of layers
required to observe the qualitative resonance phenomenon. The on-site
potential sequence of the Hamiltonian under consideration is \{$\phi ,\phi
_{2},...,\phi $\}, with the first and last on-site potentials being equal.
Since the parameters $\phi $ in the shallow circuit are related to $%
V_{1}\tau $ in the continuous-time Hamiltonian, scanning the circuit
parameter $\phi $ corresponds to scanning $V_{1}$ in the continuous-time
evolution. Therefore, the number and positions of the resonance peaks depend
only on the terms containing the phase $\phi $. The peaks originate from the
interference of different propagation paths.\ In Eq. ({\ref{P}}), $%
\sum_{i=1}^{N_{T}-1}\phi _{k_{i}}$ contributes the phase factor $\phi $ at
most $N_{T}-1$ times (from the path \{$k_{i}=1$\} or \{$k_{i}=N$\}) and at
last $0$ times (from the path \{$k_{i}\neq 1$\}). Therefore, rearranging the
expression for $P$ in powers of $e^{i\phi },$ we obtain $P_{0\cdot \cdot
\cdot 1}=\sum_{m=-(N_{T}-1)}^{(N_{T}-1)}g_{m}e^{-im\phi }.$ Noting that $%
P_{0\cdot \cdot \cdot 1}$ is real, we have $g_{m}=g_{-m}^{\ast }.$ Finally,
this yields%
\begin{equation}
P_{0\cdot \cdot \cdot 1}=\sum_{m=0}^{N_{T}-1}g_{m}\cos (m\phi ).
\end{equation}%
Therefore the number of peeks is $N_{T}-1$. In the continuous-time evolution
limit, the number of resonance peaks exhibited by a system with $N>2$ is $%
n=N-2.$ This arises because the central $N-2$ sites can be treated as a
composite subsystem possessing $N-2$\ energy levels; resonant tunneling
occurs whenever $V$ matches any of these levels. Consequently, for a shallow
quantum circuit with $N_{T}=n+1$, the number of observed peaks coincides
with that in the continuous-time limit. For the $N=2$ case, the $n=1$
resonance peak also matches the number of peaks produced by the shallow
circuit with $N_{T}=2$. Thus, for the few-qubit systems under consideration,
shallow circuits with $N_{T}=n+1$ layers exhibit the same number of
resonance peaks as in the continuous-time evolution limit. This finding
follows from the structure of the system Hamiltonian.

As shown in table. {\ref{comparison}, }the spacing between resonance peaks
observed in the shallow circuit is governed by the same factors as in the
continuous-time evolution limit. For $N=2$ and $3,$ the resonance peak
position in the continuous-time evolution depend on the on-site potential $%
V_{2}$. In the corresponding shallow circuits, the peak are observed at $%
\phi =\alpha ,$ where $\phi $ corresponds to $V_{1}$ and $\alpha $ to $%
V_{2}. $\ For $N=4,$ the peak positions depend on the on-site potential $%
V_{2}$ and the coupling $J_{2}$. The analogous behavior observed in the
shallow circuit is that the peak positions are determined by $\theta _{2}$
and $\alpha ,$ where $\theta _{2}$ corresponds to $J_{2}.$ For $N=5$, two of
three peak positions are related to $J_{2}$ and $V_{2},$ and the third one
is at $V=0.$ In the shallow quantum circuit, two of the three peaks depend
on $\theta _{2} $ and $\alpha $, and the third is fixed at $0,$ independent
of these parameters. In summary, for the systems studied here, shallow
circuits with $N_{T}=n+1$ layers capture certain features of the
continuous-time dynamics. Specifically, they reproduce both the number of
resonance peaks and the parametric dependence of their positions. We refer
to this level of correspondence as qualitative agreement, and we term the
corresponding simulation in shallow circuits a qualitative simulation of the
continuous-time dynamics.

We discuss the parameter constraints under which shallow-circuit simulations
qualitatively capture continuous-time dynamics. First, the analytical
expression of $P_{0\cdot \cdot \cdot 1}$ imposes parameter constraints for $%
N=4$ and $N=5$, while no such constraints arise for $N=2$ and $N=3$.
Specifically, for $N=4$, it requires $g_{2}>0$ and {$|t_{0}|<1$ }; for $N=5$%
, it requires {$|t_{0}|<1$}. Second, we set the edge--bulk hopping $J_{1}$
to be much smaller than the bulk hoppings, forming a tightly bound quantum
well. In the limit $J_{1}\rightarrow 0$, this well hosts $N-2$ discrete
energy levels, which depend primarily on bulk parameters. Third, the
evolution time $t$\ in the reference continuous-time dynamics is restricted
to scenarios where the particle reaches the last site without multiple
reflections, as repeated bounces would introduce additional resonances and
obscure the comparison. Given these considerations, we focus on system sizes
$N=2-5$, where analytical results for shallow circuit are accessible, and
the resonance phenomenon in the continuous-time limit is inferred from the
analysis of the Hamiltonian under a physically motivated parameter hierarchy
and further confirmed by numerical simulations. For larger $N$, the
parameter space becomes more complex and the numerical evolution may exhibit
richer behavior; whether shallow circuits remain capable of capturing the
features of the continuous-time dynamics in such regimes is therefore left
as an open question.

\subsection{CR model}

\label{CR model}
\begin{figure}[b]
\centering
\includegraphics[bb=0 0 930 923, width=10 cm, clip]{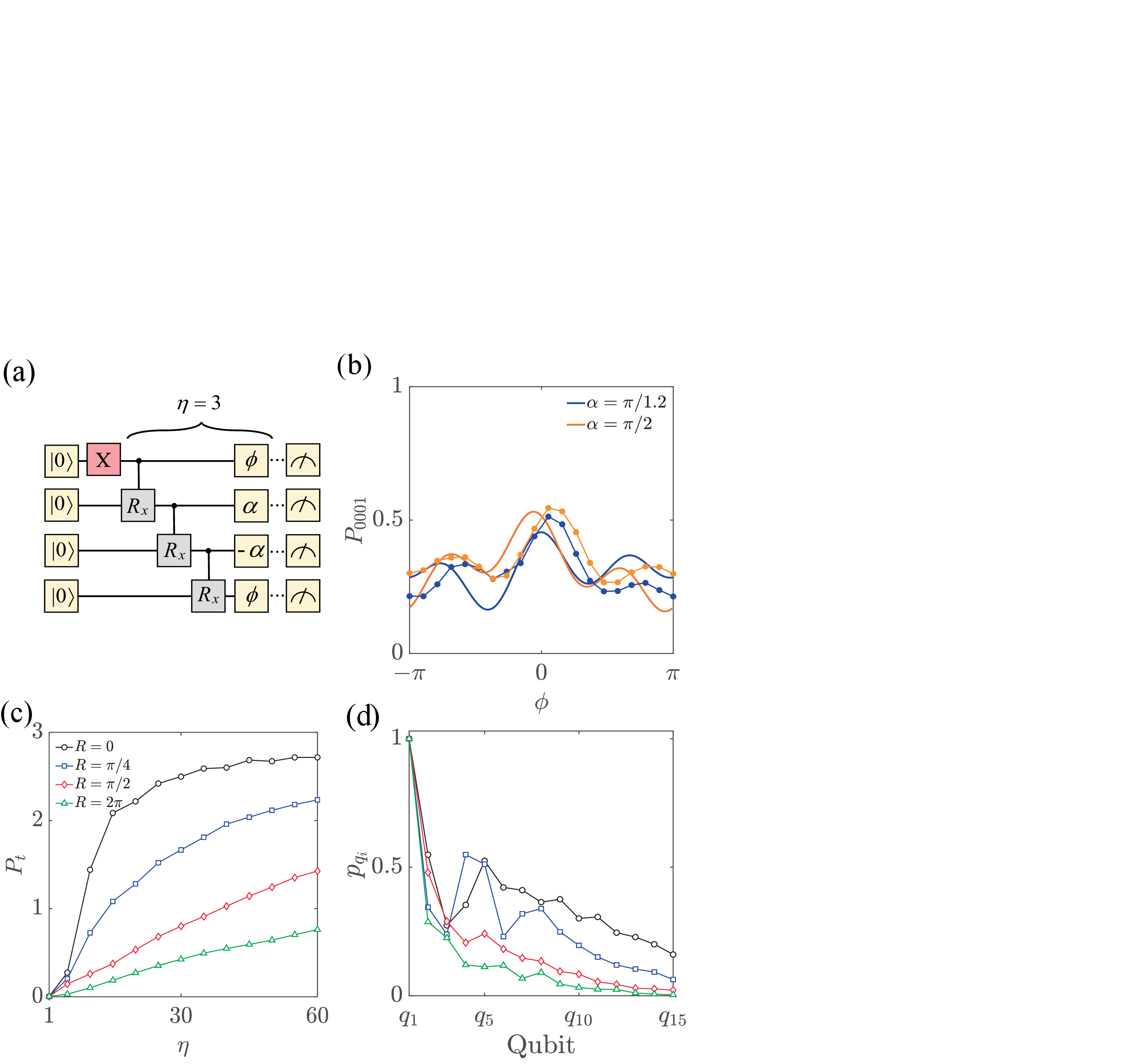}
\caption{(Color online) (a) 4-qubit circuit. The two-qubit XY gates are
replaced by controlled-$R_{x}$ gates. (b) The probability of finding the
spin excitation on the last qubit as function of $\protect\phi $. (c) The
probability of observing the spin excitation on the last third of the qubits
varies with Trotter step $\protect\eta $. (d) The probability distribution
of spin excitation at $\protect\eta =10$. (c) and (d) use the same legend
and drawing parameter $\protect\theta =\protect\phi =\protect\pi /2$.}
\label{fig3}
\end{figure}
\begin{figure*}[t]
\centering
\par
\includegraphics[bb=0 0 2362 601, width=16 cm, clip]{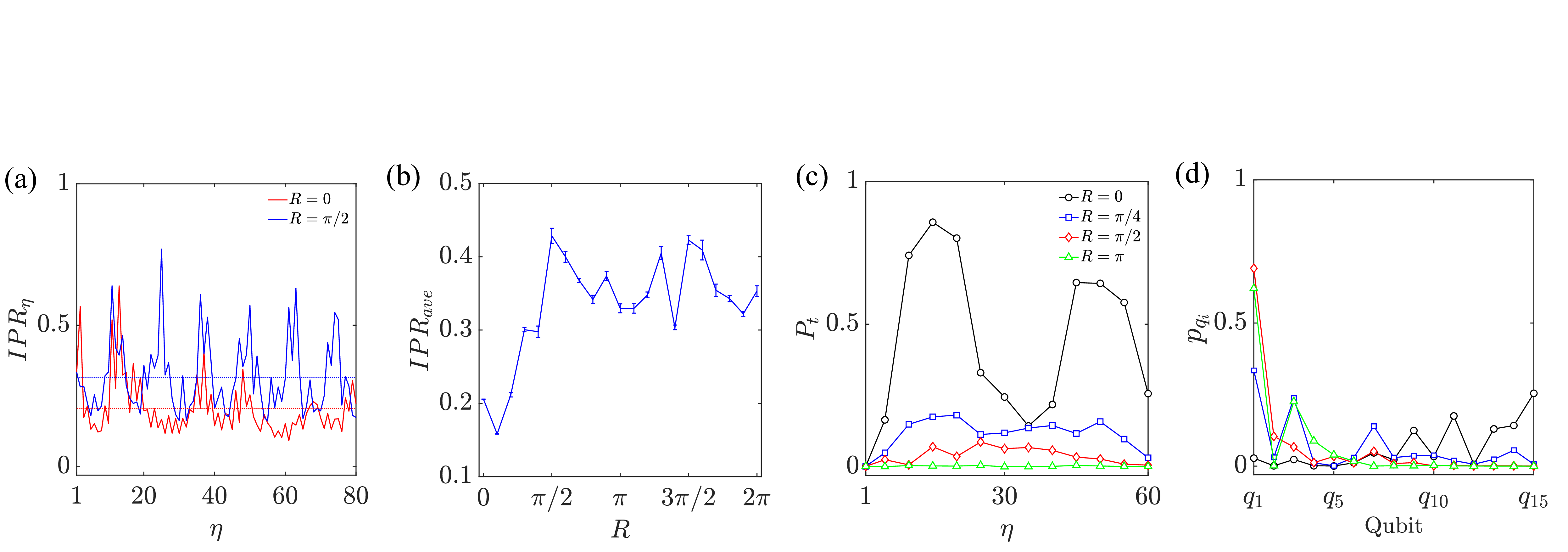}
\caption{(Color online) (a) The IPR$_{\protect\eta }$ varies with Trotter
number $\protect\eta $, the drawing parameters are $N=15,N_{T}=80,\protect%
\theta =\protect\phi =\protect\pi /2$. The red and blue lines represent the
ordered and disordered case respectively. The horizontal lines are the
average value of IPR$_{\protect\eta }$, i.e. IPR$_{ave}$. (b) IPR$_{ave}$ as
function of the degree of randomness. For a fixed $R$, we have $20$ data.
The blue line is the average value of the $20$ data, and error bar is the
variance. (c) The physical quantity $P_{t}$ is plotted in disordered
configuration. (d) The probability distribution of spin excitation when $%
\protect\eta =10$. (c) and (d) share the same legend.}
\label{fig4}
\end{figure*}

In this section, the propagation of the spin excitation through controlled-$%
R_{x}$ gates is investigated. In the previous section, the spin excitation
is created by a NOT gate and propagated through the XY gate. Here, we study
the behavior of the spin excitation propagating through the controlled-$%
R_{x} $ gates. The controlled-$R_{x}$ gate is expressed as%
\begin{equation}
U^{CR_{x}}=\left\vert 0\right\rangle \left\langle 0\right\vert \otimes
I+\left\vert 1\right\rangle \left\langle 1\right\vert \otimes e^{-i\theta
\sigma _{x}/2},
\end{equation}%
the matrix representation of $U^{CR_{x}}$ is shown in Fig. \ref{fig1}(c).
Consider the situation that a spin excitation on the first qubit passes
through a controlled-$R_{x}$ gate, we get the following equation%
\begin{equation}
U^{CR_{x}}\left\vert 10\right\rangle =\cos \frac{\theta }{2}\left\vert
10\right\rangle -i\sin \frac{\theta }{2}\left\vert 11\right\rangle .
\end{equation}%
The above equation indicates that after passing through a controlled-$R_{x}$
gate, the spin excitation becomes a two-qubit entangled state. From the view
point of propagation of the bit-flip error, this indicates that the
controlled-$R_{x}$ gate can transform a single-qubit error to a multi-qubit
error. We take the four-qubit circuit in Fig. \ref{fig3}(a) as a example to
study the behavior of spin excitation propagating through controlled-$R_{x}$
gates under the discrete-time evolution with a large step size. We observed
the probability of finding the spin excitation on the $4$th qubit and denote
the probability as $P_{0001}$. In Fig. \ref{fig3}(b), $P_{0001}$ shows two
resonant peaks, the distance between the two resonant peaks is changed as
the $\phi $ varies. This indicates that even if the spin excitations are
propagated by the controlled-$R_{x}$ gate, when the step size is large we
can qualitatively\ observe the resonance phenomenon that occurs in the
continuous-time limit.

\section{Localization for large Trotter step size}

\label{Localization for large Trotter step size}

The localization phenomenon \cite{Anderson,Wiersma,Vollhardt}
induced by disordered on-site potentials exhibits localized states, and the
localization is stronger with the increasing randomness. In this section,
we study whether the analogous localization phenomenon can be observed in
the discrete-time evolution when the Trotter step size is large.

We begin with the Hamiltonian in Eq. (\ref{TransXY})\ with $%
J_{j}=J$. The parameters in shallow quantum circuit are defined as $J\tau
=\theta ,V_{j}\tau =\phi _{j}.$ Since the on-site potentials $V_{j}$
correspond to $\phi _{j}$, thus we study the localization induced by the
disordered $U^{Z}$ gates in shallow circuit. We set one layer of parameters
for $U^{Z}$ in Eq. (\ref{TS})\ as \{$\phi _{j}$\}$\equiv $\{$\phi _{1},-\phi
_{2},\phi _{3},-\phi _{4},\cdots $\}, where $\phi _{j}=\phi +r_{j}$, $%
r_{j}\in \lbrack -R$, $R]$ is a random number. \{$\phi _{j}$\} is the same
for each layer of $U^{Z}$. \{$\phi _{j}$\} is ordered (periodic)
configuration when $R=0$ and disordered configuration when $R>0$. The
inverse participation ratio \cite{Haake} (IPR) is a measure of localization
and defined as%
\begin{equation}
IPR_{\eta }=\sum_{i=1}^{N}|P_{i}(\eta )|^{4},
\end{equation}%
where%
\begin{eqnarray}
P_{i}(\eta ) &=&\left\langle i\right. \left\vert \psi (\eta )\right\rangle ,
\nonumber \\
\left\vert \psi (\eta )\right\rangle
&=&(\prod_{j=1}^{N-1}U_{j,j+1}^{XY}(\theta
_{j})\prod_{j=1}^{N}U_{j}^{Z}(\phi _{j}))^{\eta }\left\vert \psi
(0)\right\rangle .
\end{eqnarray}%
$\left\vert \psi (\eta )\right\rangle $ represents the quantum state at the $%
\eta $th Trotter step, $P_{i}(\eta )$ represents the corresponding amplitude
at the $i$th qubit. In general, IPR$_{\eta }$ varies from $1/N$ (system
size) to $1$ and a large value of IPR$_{\eta }$ means a stronger
localization effect. The localization of $\left\vert \psi (\eta
)\right\rangle $ changes with $\eta $, thus the average IPR is introduced to
character the average level of localization during the whole discrete-time
evolution \cite{ZollerSA},
\begin{equation}
IPR_{ave}=\frac{1}{N_{T}}\sum_{\eta =1}^{N_{T}}IPR_{\eta }.
\end{equation}%
In Fig. \ref{fig4}(a), we plot $IPR_{\eta }$ (the solid lines) as function
of $\eta $ for the ordered ($R=0$) and disordered ($R=\pi /2$) configuration
respectively. The drawing parameters are $N=15,N_{T}=80,\theta
=\phi =\pi /2$. $IPR_{\eta }$ for both the ordered (the red lines) and
disordered (the red lines) configuration show a periodic-like behavior and
is larger than $1/N$, which means $\left\vert \psi (\eta )\right\rangle $
exhibits localization effect in both cases. However, the average value (the
horizontal line), i.e. $IPR_{ave}$, of the blue line is larger than the red
line, and the peaks of the blue line are closer to $1$. This indicates
stronger localization in the disordered case. Furthermore, in Fig. \ref{fig4}%
(b), we show the $IPR_{ave}$ varying with the degree $R$ of the randomness.
With the increase of disorder, the localization becomes stronger. In summary, the shallow quantum circuits exhibit localization
induced by disorder $R_{z}$ gates, with its degree increasing as disorder
strength increases. We refer to this correspondence as qualitative agreement
with the localization in the continuous-time dynamics.

In this study, the propagation of the bit-flip error (i.e. an unwanted NOT
gate) is similar to the transport of the spin excitation, thus the
localization implies that single bit-flip error propagated by disordered \{$%
U_{j}^{Z}(\phi _{j})$\} does not affect the measurement on a distant qubit.
To illustrate this point, we propose a physical quantity $P_{\mathrm{t}%
}\equiv \sum_{q_{i}=2N/3}^{N}p_{q_{i}}$ which is the average probability of
finding the spin excitation on the last third of the qubits, where $%
p_{q_{i}} $ denotes the probability of finding the error on the $i$th qubit.
The smaller the $P_{\mathrm{t}}$, the shorter the distance the spin
excitation travels. As shown in Fig. \ref{fig4}(c), $P_{\mathrm{t}}$ is
lower when $r_{i}\neq 0$, which demonstrates that only a little probability
is propagated to the last several qubits. We plot $P_{\mathrm{t}%
}$ with different randomness $R$, the results show that $P_{\mathrm{t}}$
decreases as the degree of randomness increases. In Fig \ref{fig4}(d), we
plot $p_{q_{i}}$ at $\eta =10$. As we can see, more probabilities are
propagated to the last few qubits for $r_{i}=0$ and are localized at the
first few qubits for $r_{i}\neq 0$. Figure \ref{fig4}(d) also show that the
greater the degree of randomness, the stronger the localization phenomenon. Above results indicates that the disordered parameters reduce the
influence of a bit-flip error on the first qubit on the measuremen at the
last qubit.

The above conclusion still holds if XY gates in the circuit are replaced by
controlled-$R_{x}$. As shown in Fig. \ref{fig3}(c), $P_{t}$ grows with the
increasing Trotter steps. However, $P_{\mathrm{t}}$ becomes lower when the
random perturbation is applied to a layer of $R_{z}$ gates, which means less
probabilities are propagated to the end of the circuit. With the increasing
degree of disorder, the inhibiting effect is more significant. In Fig. \ref%
{fig3}(d), we plot the probability distribution of spin excitation when $%
\eta =10,\theta =\phi =\pi /2$. These results show that, with a strong
randomness of \{$\phi _{j}$\}, the spin excitation will not be propagated to
the farther qubits. At the same time, this also shows that the randomness of
the parameters will inhibit the propagation of the bit-flip error and
protect the measurement on the distant qubits is not affected.

\section{Conclusions}

\label{Conclusions}

We study the transport of the spin excitation in the discrete-time evolution
using Trotter circuits with a large step size, and qualitatively observe the
quantum phenomena analogous to those in the continuous-time limit%
, i.e. the resonant tunneling and localization. We observe the resonance
phenomenon during the transport of the spin excitation in systems of sizes $%
N=2,3,4,5$ respectively. The probability distribution of spin excitations
propagating through several Trotter steps agree qualitatively with that in
the continuous-time limit. The corresponding minimum number of Trotter steps
is given for each system size. In a Trotter circuit with random parameters
of $R_{z}$ gates, we can qualitatively observe the localization phenomenon
of the spin excitation distribution even with a large step size. We study
the spin excitation propagating through the XY gates and also through the
controlled-$R_{x}$ gates. Our research indicates that a discrete-time
quantum simulator with a large step size can qualitatively capture some
physical phenomena in the continuous-time limit. Qualitative observations
require fewer quantum gates than quantitative calculations and therefore are
a promising application on near-term quantum computers. In quantum
computing, some errors, such as bit-flip errors, behave like spin
excitations, thus, our finding can be used to understand the propagation of
these errors in the quantum circuits.

\ack
I am grateful to Ying Li for the discussions and help in preparing the manuscript.
This work is supported by the National Natural Science Foundation of China
(Grants No. 12305018), the Fundamental Research Funds for the Central
Universities (Grant No. 2025MS080).

\section*{Appendix}

\label{Appendix}

\subsection{The amplitude of the paths}

\label{The amplitude of the paths}

The time evolution operator of transverse XY model%
\begin{equation}
H=\sum_{j}\left[ \frac{J_{j}}{4}(\sigma _{j}^{x}\sigma _{j+1}^{x}+\sigma
_{j}^{y}\sigma _{j+1}^{y})+\frac{V_{j}}{2}\sigma _{j}^{z}\right]
\end{equation}%
can be expanded approximately%
\begin{equation}
e^{-iHt}=e^{-i\sum_{j}\left[ J_{j}t(\sigma _{j}^{x}\sigma _{j+1}^{x}+\sigma
_{j}^{y}\sigma _{j+1}^{y})/4+V_{j}t\sigma _{j}^{z}/2\right] }\approx \lbrack
\prod_{j=1}^{N-1}U_{j,j+1}^{XY}(\theta _{j}\tau
)\prod_{j=1}^{N}U_{j}^{Z}(V_{j}\tau )]^{N_{T}},
\end{equation}%
where $\tau =t/N_{T},$ $\theta _{j}=${\ }$J_{j}\tau ,$ $\phi _{j}=V_{j}\tau
, $ $U_{j,j+1}^{XY}(J_{j}\tau )${$=e^{-i\theta _{j}\sigma _{j}^{x}\sigma
_{j+1}^{x}/4}e^{-i\theta _{j}\sigma _{j}^{y}\sigma _{j+1}^{y}/4}$ is a
two-qubit gate between $j$ and $j+1$, and }$U_{j}^{Z}(\phi _{j})$$=e^{-i\phi
_{j}\sigma _{j}^{z}/2}$ is a single-qubit gate. Expanding the exponential
function into trigonometric functions yields%
\begin{equation}
\left\{
\begin{array}{c}
U_{j,j+1}^{XY}(\theta _{j})=\cos ^{2}\left( \theta _{j}/4\right) -i\cos
\left( \theta _{j}/4\right) \sin \left( \theta _{j}/4\right) \left(
Y_{j}Y_{j+1}+X_{j}X_{j+1}\right) \\
-\sin ^{2}\left( \theta _{j}/4\right) X_{j}X_{j+1}Y_{j}Y_{j+1} \\
U_{j}^{Z}(\theta _{j})=\cos \left( \theta _{j}/2\right) -i\sin \left( \theta
_{j}/2\right) Z_{j}%
\end{array}%
\right. ,
\end{equation}%
it is not difficult to check that
\begin{equation}
U_{j,j+1}^{XY}\left\vert l\right\rangle =\left\{
\begin{array}{c}
\cos \left( \theta _{j}/2\right) \left\vert j\right\rangle -i\sin (\theta
_{j}/2)\left\vert j+1\right\rangle ,l=j \\
\cos \left( \theta _{j}/2\right) \left\vert j+1\right\rangle -i\sin \left(
\theta _{j}/2\right) \left\vert j\right\rangle ,l=j+1 \\
\left\vert l\right\rangle ,l\neq j,j+1%
\end{array}%
\right.  \label{U}
\end{equation}%
and%
\begin{equation}
U_{j}^{Z}X_{l}\left\vert l\right\rangle =\left\{
\begin{array}{c}
e^{i\phi _{j}/2}\left\vert j\right\rangle ,l=j \\
e^{-i\phi _{j}/2}\left\vert j\right\rangle ,l\neq j%
\end{array}%
\right. .  \label{V}
\end{equation}%
Equation ({\ref{U}}) shows that when an error propagates through an XY gate,
it can only stay on the original qubit or move to an adjacent one, and no
additional errors are generated. When the gate sequence in Eq. ({\ref{TS}})
acts on the initial state, the error propagates along several paths to reach
the final qubit, where it can be detected. Each path contributes an
amplitude, and the squared modulus of the total amplitude gives the
probability of detecting the error on the last qubit.

The amplitude of a path can be obtained according to Eq. ({\ref{U}}) and ({%
\ref{V}}). A path can be represented by the sequence \{$k_{i}$\} with $k_{i}$
denoting the qubit position at which the error resides at the end of the $i$%
-th layer. The quantum gates traversed by the path determine its
corresponding amplitude. Equation ({\ref{U}}) establishes the rules for
transmission through an $XY$ gate acting on the $j$th qubit. Specifically,
if the error is transmitted from qubit $j$ to the adjacent qubit\ upon
passing through the XY gate, the path amplitude acquires a multiplicative
factor of $-i\sin (\theta _{j}/2)$. Conversely, if the error remains on
qubit $j$ after the gate, the amplitude acquires a factor of $\cos \left(
\theta _{j}/2\right) $. Furthermore, Equation ({\ref{V}}) specifies that
when the error passes through the $i$-th layer of $R_{z}$ gates, the path
amplitude is multiplied by a phase factor $e^{-iS/2+i\phi _{k_{i}}}$ with $%
S=(N_{T}-1)\sum_{j=1}^{n}\phi _{j}$. Based on the above rules, we can
provide a general expression for the amplitude of the paths. The amplitude
of a path consists of three contributions. All two-qubit gates XY$_{j,j+1}$
along a path that involve no (resp. do) qubit transfer contribute a factor $%
\prod\limits_{j}\cos ^{c_{j}}(\theta _{j}/2)$ (resp. $\prod\limits_{j}(-i%
\sin (\theta _{j}/2))^{c_{j}}$) with $c_{j}$ counting the number of
satisfying the respective condition. The contribution of all the $R_{z}$
gate is $e^{-iS/2}e^{i\sum_{i=1}^{N_{T}-1}\phi _{k_{i}}}.$ Finally, the
amplitude of a path can be expressed as%
\begin{equation}
A(\{k_{i}\})=a(\{k_{i}\})e^{-iS/2}e^{i\sum_{i=1}^{N_{T}-1}\phi _{k_{i}}}
\end{equation}%
where $a(c_{1},c_{2})=\prod\limits_{j}\cos ^{c_{j}}(\theta
_{j}/2)\prod\limits_{l}(-i\sin (\theta _{l}/2))^{c_{l}}$ and $c_{1}$ and $%
c_{2}$ depends on the path \{$k_{i}$\}. It is straightforward to obtain that
the probability of detecting the error on the last qubit
\begin{equation}
P_{0\cdot \cdot \cdot
1}=|\sum_{\{k_{i}\}}A(\{k_{i}\})|^{2}=\sum_{\{k_{i}\},\{k_{i}^{\prime
}\}}a(\{k_{i}\})a^{\ast }(\{k_{i}^{\prime }\})e^{i\sum_{i=1}^{N_{T}-1}\phi
_{k_{i}}}e^{-i\sum_{i=1}^{N_{T}-1}\phi _{k_{i}^{\prime }}}.  \label{Papp}
\end{equation}

\subsection{Three-qubit systems}

\label{Three-qubit systems}

As a complement to the discussion in the main text, we now study the
resonance phenomenon in shallow quantum circuits for the three-qubit system.
The Hamiltonian of the transverse-field XY chain with $3$ sites reads%
\begin{equation}
H=J\sum_{j}^{2}(\sigma _{j}^{x}\sigma _{j+1}^{x}+\sigma _{j}^{y}\sigma
_{j+1}^{y})/4+V_{1}\sigma _{1}^{z}/2+V_{2}\sigma _{2}^{z}/2+V_{1}\sigma
_{3}^{z}/2.  \nonumber
\end{equation}%
The corresponding time evolution can be approximated,%
\begin{eqnarray}
U(t) &=&e^{-iHt}\approx \lbrack U_{1,2}^{XY}(J\tau )U_{2,3}^{XY}(J\tau
)U_{1}^{Z}(V_{1}\tau )\times  \nonumber \\
&&U_{2}^{Z}(V_{2}\tau )U_{3}^{Z}(V_{1}\tau )]^{N_{T}},
\end{eqnarray}%
where $U_{j,j+1}^{XY}(J_{j}\tau )=e^{-i(\sigma _{j}^{x}\sigma
_{j+1}^{x}+\sigma _{j}^{y}\sigma _{j+1}^{y})J_{j}\tau /4}$, $U_{j}^{Z}\left(
V_{j}\tau \right) =e^{-i\sigma _{1}^{z}V_{j}\tau /2}$, $\tau =t/N_{T}$. The
right side of above equation{\ represents the discrete-time evolution. }The
quantum circuit implementing the discrete-time evolution is shown in Fig.%
\textit{\ }\ref{fig2}(b1). We concern the time evolution of single spin
excitation. When the single spin excitation can be treated as a particle,
the transverse-field XY chain is equal to a $3$-site tight-binding chain
[see Fig. \ref{fig2}(b3)],%
\begin{eqnarray}
H &=&J\left( \left\vert 100\right\rangle \left\langle 010\right\vert
+\left\vert 010\right\rangle \left\langle 100\right\vert +\mathrm{h.c.}%
\right) +V_{1}\left\vert 100\right\rangle \left\langle 100\right\vert
\nonumber \\
&&+V_{2}\left\vert 010\right\rangle \left\langle 010\right\vert
+V_{1}\left\vert 001\right\rangle \left\langle 001\right\vert .
\label{HofTripleWell}
\end{eqnarray}%
where $J$ is the coupling strength, the potentials on three sites are $%
V_{1},V_{2}$ and $V_{1}$ respectively. In the case that $V_{1}$ is the only
variable parameter, the resonance phenomenon means that the probability, $%
P_{3}(t,V_{1})$, of finding particle on the $3$th site reaches maximum at $%
V_{1}=V_{2}$. We numerically simulate the discrete-time evolution with a
large $N_{T}$ and show the resonance phenomenon. The initial state is $%
\left\vert 100\right\rangle $. In Fig. \ref{fig2}(b4),\ we plot $P_{3}$ as
the function of $V_{1}$ at $t=22$ (units of $1/J$,{\ }$J=0.1$) with the
transparent lines. $P_{3}$ has one resonance peak at $V_{1}=V_{2}$.
Similarly, we can consider the resonance phenomenon with a triple-well
system [see Fig. \ref{fig2}(b2)], whose Hamiltonian can be written as Eq. (%
\ref{HofTripleWell}). In the triple-well system, the probability of the
particle tunneling from the left well to the right well reaches maximization
at $V_{1}=V_{2},$ when the resonance occurs.

As for the small $N_{T}$, we find that only two Trotter steps are required
for the $3$-qubit circuit to exhibit resonant tunneling. The parameters of
the $3$-qubit circuit are redefined as $J\tau =\theta ,V_{1}\tau =\phi
,V_{2}\tau =\alpha $. As shown the blue, red and purple dashed lines in Fig. %
\ref{fig2}(b1), the spin excitation goes through three paths. The blue and
red paths contribute $\cos (\theta /2)(-i\sin (\theta /2))(-i\sin (\theta
/2))e^{-i\alpha /2}$ to the amplitude, and the purple path contributes $%
-i\sin (\theta /2)\cos (\theta /2)e^{i\left( \alpha /2-\phi \right) }\cos
(\theta /2)(-i\sin (\theta /2))$ to the amplitude. The amplitude of the
final state on the third qubit is%
\begin{equation}
A_{001}=-\sin ^{2}(\theta /2)\cos (\theta /2)\left( 2e^{-i\alpha /2}+\cos
(\theta /2)e^{i\left( \alpha /2-\phi \right) }\right) .
\end{equation}%
Accordingly, the probability of finding the spin excitation on the $3$rd
qubit is%
\begin{equation}
P_{001}(\theta ,\phi )=\sin ^{4}(\theta /2)\cos ^{2}(\theta /2)(4+\cos
^{2}(\theta /2)+4\cos (\theta /2)\cos \left( \alpha -\phi \right) ).
\label{P001}
\end{equation}%
The interference term $4\sin ^{4}(\theta /2)\cos ^{3}(\theta /2)\cos \left(
\alpha -\phi \right) $ dominates the resonant tunneling effect. %
From the above results, the resonance observed in three-qubit circuits with
two Trotter steps exhibits two characteristics that are consistent with
those in the continuous-time limit. First, both cases exhibit a single peak.
Second, the peak is located at $\phi =\alpha ,$ mirroring the behavior
observed in the continuous-time limit where the resonance occurs at $%
V_{2}=J_{2}.$ In Fig. \ref{fig2}(b4), we plot $P_{001}$ as function of $%
\phi $ with $\theta =\pi /2$ and $\alpha =0,-\pi /2$. The accurate results
(solid lines) computed using QuESTlink coincide with the experimental
outcomes (solid lines with point symbols) computed using the IBM quantum
device "ibmq\_rome". As we can see, $P_{3}$ coincides well with $P_{001}$,
both of them have only one peak and the position of the peak is $\alpha
=\phi $ (i.e. $V_{1}=V_{2}$). The above analysis indicates that only two
Trotter steps are required for the $3$-qubit circuit to exhibit similar
resonant tunneling in the continuous-time limit.

\subsection{\textit{Four}-qubit systems}

\label{Four-qubit systems}

As a supplement to the main text, this section investigates the resonance
phenomenon in shallow quantum circuits for the four-qubit case. The
Hamiltonian of the $4$-site transverse-field XY chain we studied is
\begin{eqnarray}
H &=&J_{1}\sum_{j=1,3}(\sigma _{j}^{x}\sigma _{j+1}^{x}+\sigma
_{j}^{y}\sigma _{j+1}^{y})/4+J_{2}(\sigma _{2}^{x}\sigma _{3}^{x}+\sigma
_{2}^{y}\sigma _{3}^{y})/4  \nonumber \\
&&+V_{1}\sigma _{1}^{z}/2+V_{2}\sigma _{2}^{z}/2-V_{2}\sigma
_{3}^{z}/2+V_{1}\sigma _{4}^{z}/2.
\end{eqnarray}%
The corresponding discrete-time evolution is in the form%
\begin{eqnarray}
U(t) &=&e^{-iHt}\approx \lbrack U_{1,2}^{XY}(J_{1}\tau
)U_{2,3}^{XY}(J_{2}\tau )U_{3,4}^{XY}(J_{1}\tau )\times  \nonumber \\
&&U_{1}^{Z}(V_{1}\tau )U_{2}^{Z}(V_{2}\tau )U_{3}^{Z}(-V_{2}\tau
)U_{3}^{Z}(V_{1}\tau )]^{N_{T}},
\end{eqnarray}%
where $U_{j,j+1}^{XY}(J_{j}\tau )=e^{-i(\sigma _{j}^{x}\sigma
_{j+1}^{x}+\sigma _{j}^{y}\sigma _{j+1}^{y})J_{j}\tau }$, $U_{j}^{Z}\left(
V_{j}\tau \right) =e^{-i\sigma _{1}^{z}V_{j}\tau /2}$, $\tau =t/N_{T}$. We
plot the quantum circuit implementing the discrete-time evolution in Fig.%
\textit{\ }\ref{fig2}(c1). In the single-particle subspace, the equivalent $%
4 $-site tight-binding chain is%
\begin{eqnarray}
H &=&J_{1}\left( \left\vert 1000\right\rangle \left\langle 0100\right\vert
+\left\vert 0010\right\rangle \left\langle 0001\right\vert \right)
+J_{2}(\left\vert 0100\right\rangle \left\langle 0010\right\vert  \nonumber
\\
&&+\left\vert 0010\right\rangle \left\langle 0100\right\vert
)+V_{1}\left\vert 1000\right\rangle \left\langle 1000\right\vert
+V_{2}\left\vert 0100\right\rangle \left\langle 0100\right\vert  \nonumber \\
&&-V_{2}\left\vert 0010\right\rangle \left\langle 0010\right\vert
+V_{1}\left\vert 0001\right\rangle \left\langle 0001\right\vert .  \label{H4}
\end{eqnarray}%
$\allowbreak \allowbreak $As we marked in Fig. \ref{fig2}(c4), the coupling
strengths between neighboring sites are $J_{1},J_{2}$, and $J_{1}$
respectively, and the on-site potentials on the four sites are $%
V_{1},V_{2},-V_{2},V_{1}$ respectively. In the condition of $J_{1}\ll
J_{2},V_{2}$, we numerically simulate the discrete-time evolution of one
particle with a large $N_{T}$. In Fig. \ref{fig2}(c4), $P_{4}(V_{1},t)$,
which is the probability of finding the particle on the $4$th site, is
plotted as function of $V_{1}$ with the transparent lines. The cases of $%
V_{2}=10$ and $V_{2}=20$ are studied when $J_{1}=1$, $t=3$ (units of $%
1/J_{1} $)$,$ $J_{2}=20$. As we can see, the resonant peaks can be observed
near $\sqrt{J_{2}^{2}+V_{2}^{2}}$ and $-\sqrt{J_{2}^{2}+V_{2}^{2}}$, and the
distance between the resonance peaks varies when $V_{2},-V_{2}$ change. One
can observe the same resonance phenomenon in a triple-well system [see Fig. %
\ref{fig2}(c2)]. The energy levels of the left and right wells are $V_{1}$,
the middle well has two energy levels: $\sqrt{J_{2}^{2}+V_{2}^{2}}$ and $-%
\sqrt{J_{2}^{2}+V_{2}^{2}}$. If there is a particle in the left well at the
initial moment, then we can detect this particle in the right well with a
certain probability. When $V_{1}$ is close to $\sqrt{J_{2}^{2}+V_{2}^{2}}$
or $-\sqrt{J_{2}^{2}+V_{2}^{2}}$, the probability of finding particle in
right well reaches the maximum. The distance of the two peaks varies with $%
V_{2}$.

When $N_{T}$ is small, we find that only three Trotter steps are required
for the $4$-qubit circuit to qualitatively exhibit resonant tunneling. According to Eq. ({\ref{Papp}}), the amplitude of the final
state on the third qubit is%
\begin{equation}
A_{0001}=\sum_{m=0}^{2}f_{m}e^{-mi\phi }
\end{equation}%
where
\begin{equation}
\left\{
\begin{array}{c}
f_{2}=\frac{1}{4}\sin ^{2}\theta _{1}\sin (\theta _{2}/2)(\cos 2\alpha
+2\cos ^{2}\alpha \cos \theta _{2}) \\
f_{1}=\frac{1}{2}\sin ^{2}(\theta _{1}/2)\sin \theta _{2}\cos \alpha
(1+3\cos (\theta _{1})) \\
f_{0}=\frac{3}{4}\sin ^{2}\theta _{1}\sin (\theta _{2}/2)%
\end{array}%
\right. ,
\end{equation}%
the probability of finding the spin excitation on the $4$th qubit is%
\begin{equation}
P_{0001}=\sum_{m=0}^{2}g_{m}\cos m\phi  \label{P0001}
\end{equation}%
where $g_{0}=\sum_{i=0}^{2}f_{i}^{2},g_{1}=2f_{1}\left( f_{2}+f_{0}\right)
,g_{2}=2f_{2}f_{0}.$ Applying trigonometric transformations together with
polynomial theorems shows that the peak positions occur at $\phi =\pm
\arccos (t_{0})$ when $g_{2}<0$, where $t_{0}=-g_{1}/4g_{2}$$.$ That is,
under the condition that $g_{2}<0$ and $\left\vert t_{0}\right\vert <1$, the
resonance observed in four-qubit circuits with three Trotter steps exhibits
two characteristics that are consistent with those in the continuous-time
limit. First, both cases exhibit two peaks. Second, the peaks are located at
$\phi =\pm \arccos (-g_{1}/4g_{2}),$ mirroring the behavior observed in the
continuous-time limit where the resonance occurs at $V_{2}=\pm \sqrt{%
J_{2}^{2}+V_{2}^{2}}.$ We plot $P_{0001}$ (the solid lines) in Fig. \ref%
{fig2}(c1). The parameters are redefined as $J_{1}\tau =\theta
_{1},J_{2}\tau =\theta _{2},V_{1}\tau =\phi ,V_{2}\tau =\alpha $. With $%
\theta _{1}=\theta _{2}=\pi /1.5$, we plot two cases of $\alpha =\pi /4$ and
$\alpha =-\pi /1.5$. The distance between the peaks is changed when $\alpha $
is adjusted, which is the characteristic of the resonant tunneling effect.

\end{document}